\documentclass[reprint,aps,prl,superscriptaddress,amsmath,amssymb]{revtex4-2}

\usepackage{amsmath}
\usepackage{amssymb}
\usepackage{graphicx}
\usepackage{color}
\usepackage{hyperref}
\begin{document} 

\title{First Directional Dark Matter Limits from the MIMAC \texorpdfstring{$\mu$}{mu}-TPC Detector}

\author{Ilias Ourahou}
\author{Daniel Santos}
\author{Olivier Guillaudin} 
\author{Jean-François Muraz}
\author{Nadine Sauzet}
\affiliation{Laboratoire de Physique Subatomique et de Cosmologie (LPSC), Universit\'e Grenoble Alpes, CNRS/IN2P3, Grenoble, France}

\author{Charling Tao}

\affiliation{Centre de Physique des Particules de Marseille, IN2P3/CNRS and Université Aix-Marseille, Marseille, France \\ 
\vspace{0.4cm} 
\textnormal{\small E-mail: \textcolor{blue}{ilias.ourahou@lpsc.in2p3.fr}, \textcolor{blue}{daniel.santos@lpsc.in2p3.fr}}}


\begin{abstract}
We present a directional dark matter search using the MIMAC (MIcro-tpc MAtrix of Chambers) detector at the Modane Underground Laboratory. Operating a low-pressure gas mixture of 50\% $\text{i-C}_4\text{H}_{10}$ and 50\% $\text{CHF}_3$ at 30 mbar, without any shielding, the detector allows the reconstruction of the 3D tracks of nuclear recoils produced in the 6 liter active volume. Directionality is a crucial tool to discriminate true WIMP signals from those of background producing identical recoils, such as neutrons. We analyze data from two independent chambers with effective exposures of 495.1 and 354.2 days respectively. To estimate the background directly from the data, we apply a standard ON/OFF spatial analysis. By projecting the reconstructed tracks onto the galactic map and applying a moving kinematic energy cut depending on the WIMP mass, we compare the event rate in the signal direction with multiple OFF-source regions. Finding no significant excess of events in the direction of the Cygnus constellation, we set an upper limit on the spin-dependent WIMP-proton cross section of $\sigma_{SD-p} < 6.65 \times 10^{-36} \text{ cm}^2$ at 90\% confidence level.
\end{abstract}

\maketitle


The overwhelming cosmological and astrophysical evidence points to the existence of Dark Matter (DM), which constitutes approximately 26\% of the energy density and 86\% of the matter of the Universe\cite{REF_Planck_Cosmology}. Weakly Interacting Massive Particles (WIMPs) remain among the most compelling DM candidates \cite{REF_WIMP_Review}. Direct detection experiments look for the small amount of energy deposited when a WIMP collides elastically with an atomic nucleus in the active volume of a detector \cite{REF_Direct_Detection_General}. 

As massive non-directional experiments reach extreme sensitivities, they are approaching  the "neutrino floor" (or neutrino fog)---an irreducible background where coherent elastic neutrino-nucleus scattering (CEvNS) perfectly mimics WIMP-induced nuclear recoils \cite{REF_Billard_NeutrinoFloor}.

Directional detection offers a definitive physical signature to overcome this limitation \cite{REF_Grothaus_NeutrinoBound}. As first pointed out by Spergel in 1988 \cite{REF_Cygnus_Wind_Theory}, the Solar System's orbital motion through the dark matter halo creates an apparent WIMP "wind" from the Cygnus constellation direction. Consequently, true WIMP-induced recoils will exhibit a strong forward-backward dipole asymmetry. Measuring the 3D direction of these recoils allows us to discriminate the galactic halo signals from all other backgrounds, which possess isotropic or geometry-dependent distributions. This fundamental advantage places low-pressure gaseous Time Projection Chambers (TPCs) like MIMAC as the future for direct dark matter searches.

Importantly, phenomenological studies have demonstrated that directional detection is the only physical method capable of penetrating this barrier and searching beyond the neutrino bound \cite{REF_Grothaus_NeutrinoBound}. 

The MIMAC (MIcro-tpc MAtrix of Chambers) project is a gaseous detector designed to reconstruct the 3D tracks of low-energy nuclear recoils \cite{REF_MIMAC_Detector}. Located at the Laboratoire Souterrain de Modane (LSM), it uses a low-pressure (30 mbar) gas mixture of isobutane ($\text{i-C}_4\text{H}_{10}$) and 50\% of $\text{CHF}_3$. This gives sensitivity to spin-dependent interactions (via $^{1}$H and $^{19}$F), while the low pressure makes the tracks long enough to be measured by a pixelated anode read every 20 ns \cite{REF_MIMAC_Performance}.

In this Letter, we present the directional search for WIMPs using two independent MIMAC chambers with a common cathode (ch0 and ch1). Using two separate chambers helps to better understand the systematics and confirm the stability of the detector. Due to the different technologies applied to build the two Micromegas (Micro-MEsh Gaseous Structure) and maintenance schedules, the two chambers have different effective exposures: 495.1 days for ch0 and 354.2 days for ch1.

To accurately estimate the background without relying on Monte Carlo simulations, we use a data-driven ON/OFF spatial analysis \cite{REF_ON_OFF_Method_Astronomy}. We project the 3D axes of the measured recoils onto the galactic map. Then, we apply a moving kinematic cut: the maximum energy allowed for an event changes dynamically depending on the WIMP mass we are testing \cite{REF_SD_CrossSection_Targets}. This prevents us from including background events at energies where a WIMP could not physically produce a signal. By comparing the ON region (towards Cygnus) with several OFF regions (empty sky), we can estimate the background directly from the data. This Letter describes this methodology and the resulting exclusion limits on the WIMP-proton spin-dependent cross section.


The MIMAC detector is a micro-Time Projection Chamber ($\mu$-TPC) designed to measure both the kinetic energy and the 3D track of a recoiling nucleus. Inside the chamber, an incident particle scatters off a target gas nucleus, creating a primary ionization track, defined by a number of electron-ion pairs. Under a uniform electric field, these primary electrons drift towards the mesh of a Micromegas with a 512 $\mu m$ gap coupled to a pixelated anode \cite{REF_Micromegas_Tech}. The X and Y coordinates of the track are directly given by the activated pixels on the anode, while the Z coordinates are reconstructed using the drift velocity and a fast 50 MHz (20 ns) electronic sampling \cite{Bourrion}. 

By operating at a low pressure of 30 mbar, the track length of the recoiling nuclei with energies below 6 keV is on the order of a few millimeters. This allows the pixelated anode (with a 424 $\mu$m pitch) to accurately resolve the 3D shape and direction of the track \cite{REF_MIMAC_3D}. The gas mixture of 50\% isobutane ($\text{i-C}_4\text{H}_{10}$) and 50\% $\text{CHF}_3$ provides rich targets of $^{1}$H, $^{12}$C, and $^{19}$F nuclei, with a drift velocity of 11.5 $\mu m/ns$. However, for low-mass WIMPs, the kinematic energy transfer strongly favors the lightest target. Due to the kinematics of elastic scattering and the strong ionization quenching of heavier ions at low energies, the observable recoils in our Region of Interest (ROI) of ionization energies below 10 keV, are entirely dominated by Hydrogen. Consequently, the contributions from Carbon and Fluorine are kinematically suppressed and safely neglected, allowing us to focus exclusively on the spin-dependent WIMP-proton cross section ($\sigma_{SD-p}$).

\begin{figure}[htbp]
    \centering
    \includegraphics[width=0.9\columnwidth]{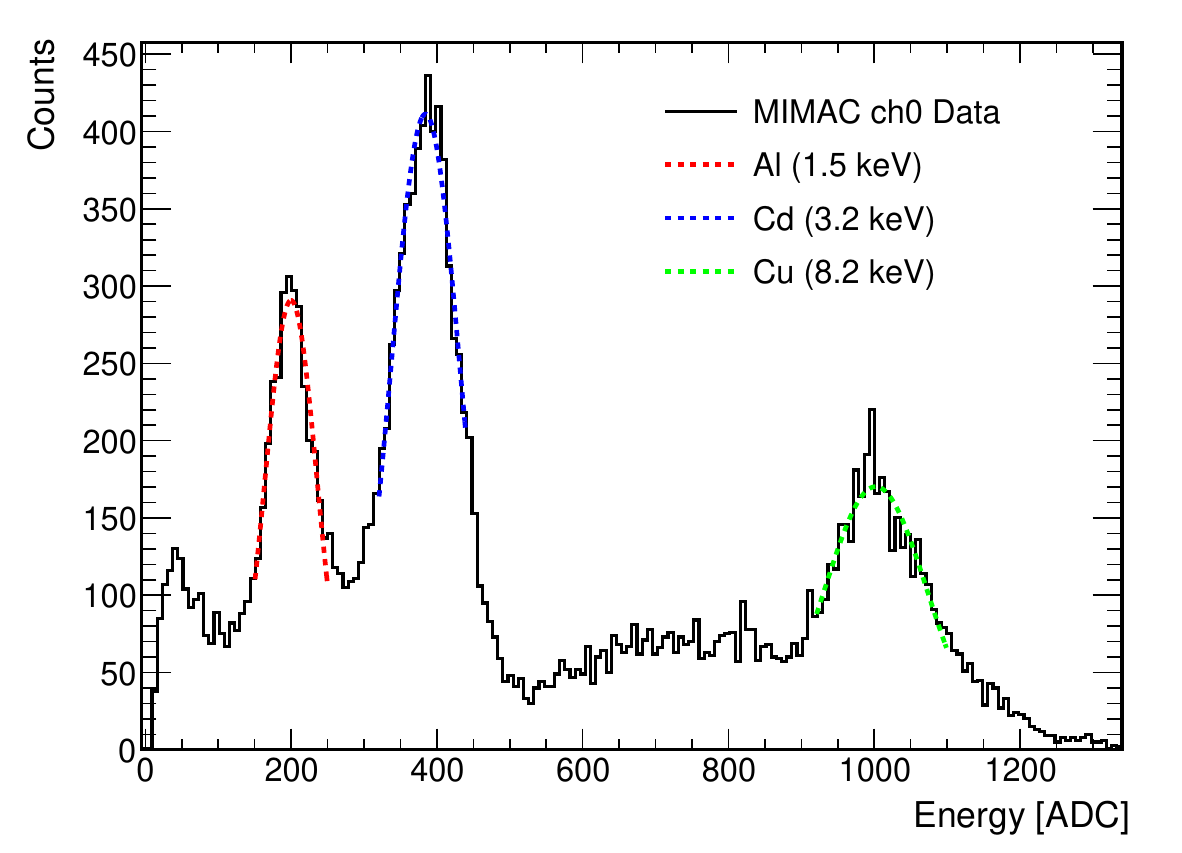} 
    \caption{Ionization energy calibration spectrum for the MIMAC detector (ch0). The distribution of the collected charge (black solid line) exhibits three distinct X-ray fluorescence peaks corresponding to Aluminium (1.5 keV), Cadmium (3.2 keV), and Copper (8.2 keV). The low-energy peak on the far left (below 100 ADC) represents the Carbon K$\alpha$ line (0.27 keV) produced by X-rays exciting the Carbon nuclei of the gas mixture. Dashed colored lines represent independent Gaussian fits used to extract the ADC centroid. This precise spectroscopic calibration establishes the linear baseline required to convert raw ADC channels into observable ionization energy (keVee). While the standard physics analysis threshold is evaluated at approximately 0.45 keVee to ensure noise-free operation, this sub-threshold Carbon peak is resolved during calibration runs.}
    \label{fig:calibration_ch0}
\end{figure}

To accurately measure the ionization energy of these events, regular calibrations were performed during data collection at the Modane Underground Laboratory. The ionization energy scale (in keVee) is calibrated (as shown in Fig.~\ref{fig:calibration_ch0}) using known X-ray fluorescence lines from metallic foils (such as Al, Cd, and Cu) placed in front of an X-generator inside the modulus chamber \cite{REF_Beaufort_HeadTail,REF_Ourahou_Deuteron_2026}. 

To avoid signal degradation caused by electron-attaching impurities (such as oxygen out-gassing or moisture), the gas mixture was circulated in a closed-loop system. This system incorporated oxygen and moisture filters, a pressure regulator, a flow controller, and a buffer volume, ensuring the chamber gas was renewed hourly. In addition, the experimental room was maintained at a controlled temperature and humidity, which contributed to the stable operation of the experiment.
The quality of the gas was regularly verified by the X-ray calibrations described above. This continuous monitoring allowed us to map any minor energy response drift, ensuring a highly stable and well-calibrated energy scale over the multi-year run.

The WIMP does not interact with electrons; it scatters off a nucleus \cite{REF_Directional_Review}. A recoiling nucleus produces less ionization than an electron of the same kinetic energy. This energy loss is described by the Ionization Quenching Factor (IQF). To convert the kinematic nuclear recoil energy (in keVnr) into the observable ionization energy (in keVee) measured by MIMAC, we use IQFs measured by COMIMAC  \cite{Muraz2016} and a parameterized quenching model to extrapolate to the lowest kinetic energy \cite{REF_Hill_Quenching_Model, REF_MIMAC_IQF_Measurement}. Properly accounting for this quenching is critical for our analysis, as it directly dictates the maximum observable energy for a WIMP of a given mass.

\begin{figure*}[t]
    \centering
    \includegraphics[width=2.0\columnwidth]{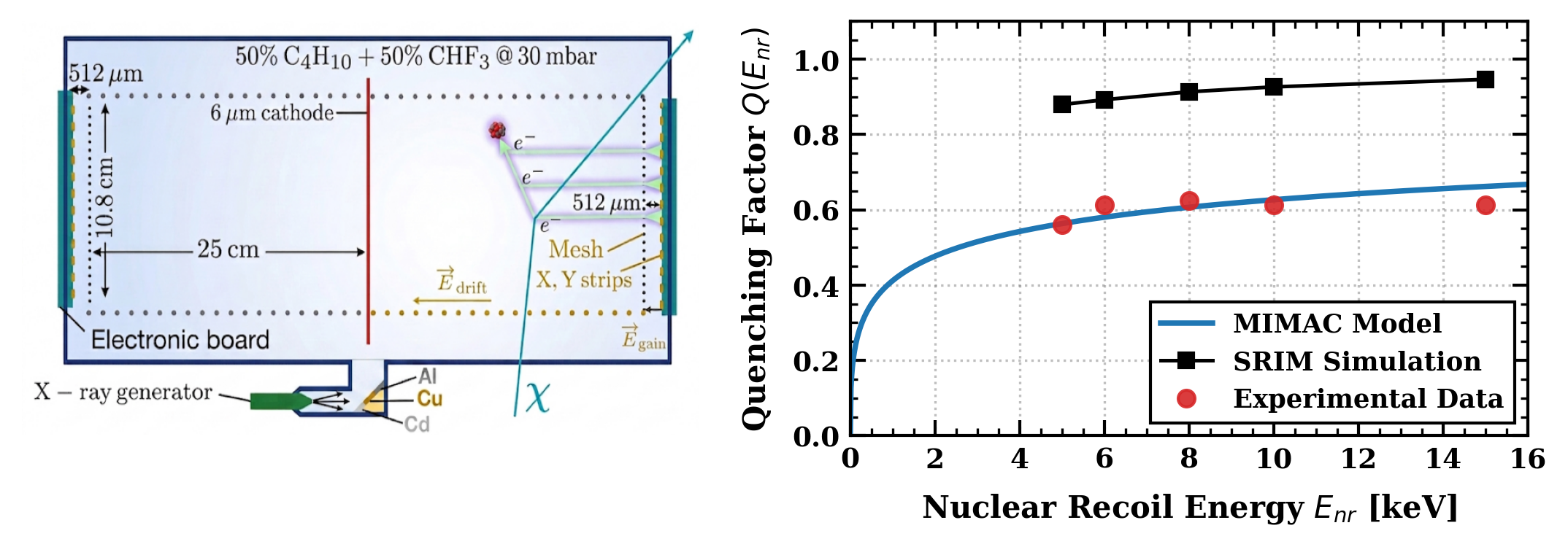} 
    \caption{(left) Schematic diagram of a bi-chamber modulus MIMAC detector, highlighting the 25~cm drift region and the X-ray calibration generator coupled with three metal foils (Al, Cd, and Cu) at one side of the modulus chamber. (right) Proton Ionization Quenching Factor (IQF) in the MIMAC gas mixture  as a function of the nuclear recoil kinetic energy. Red circles represent the experimental data points measured with the COMIMAC facility \cite{Muraz2016}, whereas the black squares connected by a solid line correspond to the SRIM simulation predictions. A discrepancy of approximately 30\% is observed between the experimental measurements and the SRIM simulation, which systematically overestimates the quenching factor. The solid blue line displays the Lindhard-based analytical parameterization: $Q(E_{nr}) = A E_{nr}^C / (B + E_{nr}^C)$, with the three fitted coefficients $A = 1.0$, $B = 1.43 \pm 0.11$, and $C = 0.38 \pm 0.04$ by Beaufort et al. \cite{Beaufort2024}. This same validated model is used in this work to extrapolate the quenching factor down to the lowest sub-keV energy detected.}
    \label{fig:quenching_comparison}
\end{figure*}


Once an ionization greater than the threshold is collected by the charge preamplifier on the grid, the acquisition is triggered by a new event, its 3D track is reconstructed using the spatial coordinates (X, Y) from the pixelated anode and the depth coordinates (Z) derived from the 50 MHz sampling and the drift velocity. To determine the principal 3D axis of the track, we apply a Principal Component Analysis (PCA) \cite{REF_Jolliffe_PCA}. However, a principal axis alone lacks the track direction. The absolute vector sense is obtained by measuring the charge asymmetry along the track, known as the Head-Tail effect. Following the method described in \cite{REF_Beaufort_HeadTail}, we process the Flash signal taking away the slow ionic contribution to extract the primary electron current profile. This profile reflects the total ionization deposition of $\mathrm{d}E/\mathrm{d}x$ energy along the drift axis (Z-axis), allowing us to recognize the Head-Tail asymmetry and assign a definitive 3D vector to each event. Crucially, the validity of both our 3D angular reconstruction and our ionization quenching factor model has been experimentally demonstrated in Ref.~\cite{REF_Beaufort_HeadTail} using monochromatic neutron fields by measuring the nuclear recoil ionization energies and angles produced by a neutron field generated at the target by a nuclear reaction. The incident neutron energies at 8~keV and 27~keV were successfully reconstructed with an angular resolution of $15^\circ$. This precise spectroscopic reconstruction of low-energy neutron fields directly validates the accuracy of our angular calculations and quenching factor measurements.

The primary challenge in direct detection is the overwhelming background produced by muons, electrons, gamma rays, and alpha particles. But the high 3D-spatial resolution of the MIMAC $\mu$-TPC allows us to exploit the morphological differences between different particle tracks. A low-energy electron undergoes a random trajectory through the gas, leaving a sparse track composed of multiple spatially separated clusters, whereas a recoiling nucleus produces a dense, and compact track. 

To quantify these morphological differences, we use a set of topological and temporal observables for each event. Spatial variables include the density of the track core (\texttt{DensityCenter}), the proportion of empty pixels along the path (\texttt{HoleDensity}), and the ratio of track length to ionization energy (\texttt{RatioSLEnergy}). These are complemented by temporal variables extracted from the Flash-ADC, such as the duration of the primary electron cloud (\texttt{PrimaryCloudDur}) and the rise-to-fall time asymmetry (\texttt{AsymFactor}). 

\begin{figure}[htbp]
    \centering
    \includegraphics[width=1.0\columnwidth]{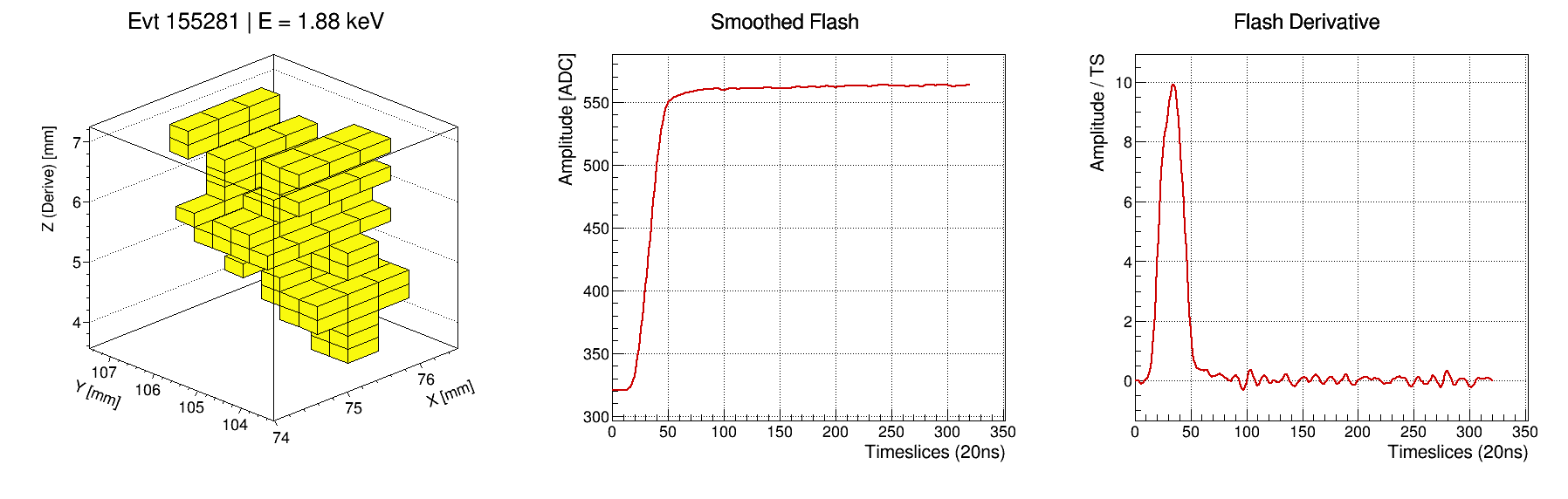}
    \includegraphics[width=1.0\columnwidth]{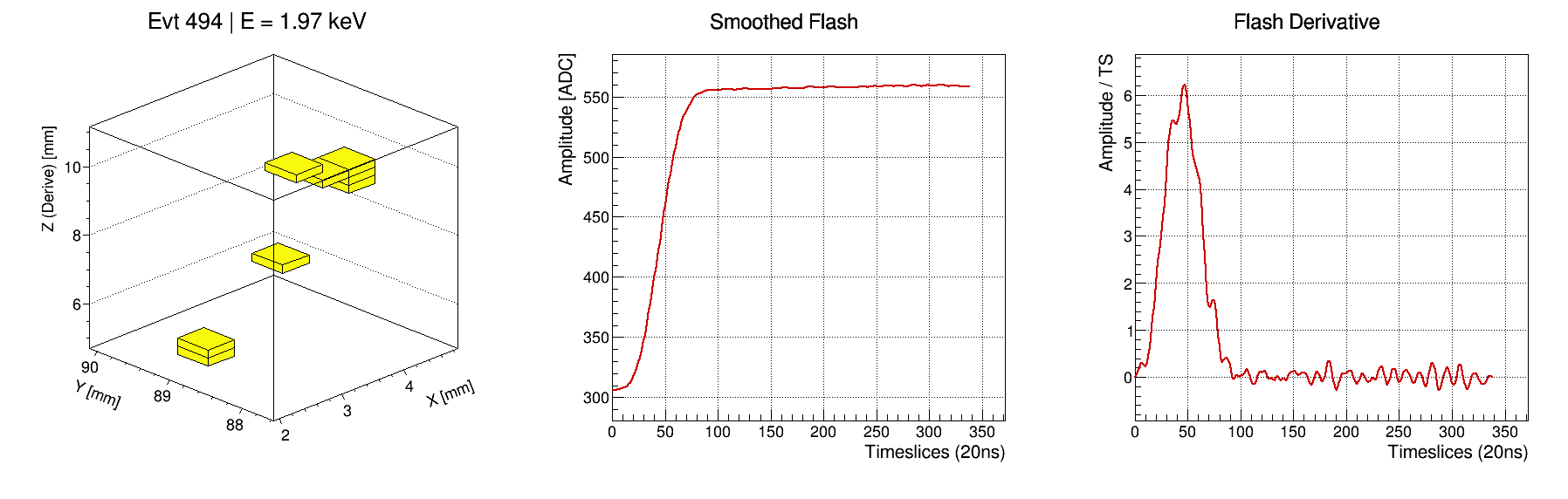}
    \caption{Topological and time profile discrimination in the MIMAC $\mu$-TPC at low energy. Top panels show a typical nuclear recoil candidate ($E = 1.88$~keVee), while bottom panels display an electron background event at a comparable energy ($E = 1.97$~keVee). Despite the similar energy deposition, their morphological signatures differ drastically. The nuclear recoil produces a compact, continuous 3D track with a high pixel density (left), corresponding to a smooth ionization profile and a single, well-defined peak in the Flash-ADC derivative (right).}
    \label{fig:comparaison}
\end{figure}
To maximize background rejection, these observables are inserted into a multivariate classifier using a Boosted Decision Tree (BDT) algorithm. To avoid systematic biases associated with pure Monte Carlo training, we employ a data-driven training strategy. The electronic background training sample is built from dedicated X-ray calibration runs. Conversely, the nuclear recoil signal training sample is extracted directly from the measurement run itself, the electronic recoil undergoes significant straggling, resulting in a clustered track with multiple spatial gaps, which translates into a noisy, multi-peaked Flash derivative. By applying highly restrictive manual topological pre-cuts—specifically bounding the 3D track length and transverse width, requiring a high pixel density, and limiting the number of peaks on the Flash derivative to select single, compact interactions(as illustrated in Fig.~\ref{fig:comparaison}—we achieve an electronic recoil rejection power of $10^5$. This implies that the probability for an electron to pass these cuts is only $10^{-5}$, allowing us to isolate a high-purity sample of nuclear recoils~\cite{REF_Riffard:2016jinst}. The rejection power of these observables has been validated in recent thermal and fast neutron capture recoil measurements  \cite{REF_Ourahou_Deuteron_2026}.

This strategy allows the BDT to learn the topology of the measured recoils, providing robust discrimination between WIMP candidates and electronic backgrounds. Only events passing an optimized BDT selection threshold are kept as valid nuclear recoil candidates for subsequent spatial ON/OFF analysis.


To perform a directional dark matter search, the reconstructed 3D tracks must be projected from the local detector frame onto the galactic map. This requires a rigorous three-step geometric transformation. First, the track is projected into the Laboratory frame using the detector's precise topographic azimuth ($\psi$) and zenith inclination ($\alpha_Z$). Second, to account for the Earth's rotation, we apply the transformation to the Equatorial celestial frame (J2000) ~\cite{Bozorgnia2011} using the exact timestamp of each event and the local coordinates of the LSM to freeze the Earth's rotation. Finally, a standard static rotation matrix (IAU 1958) is applied to transform to the Galactic map, yielding the galactic longitude ($l$) and latitude ($b$) of each event~\cite{Bozorgnia2011, Liu2011}.

\begin{figure}[htbp]
    \centering

    \includegraphics[width=1.0\columnwidth]{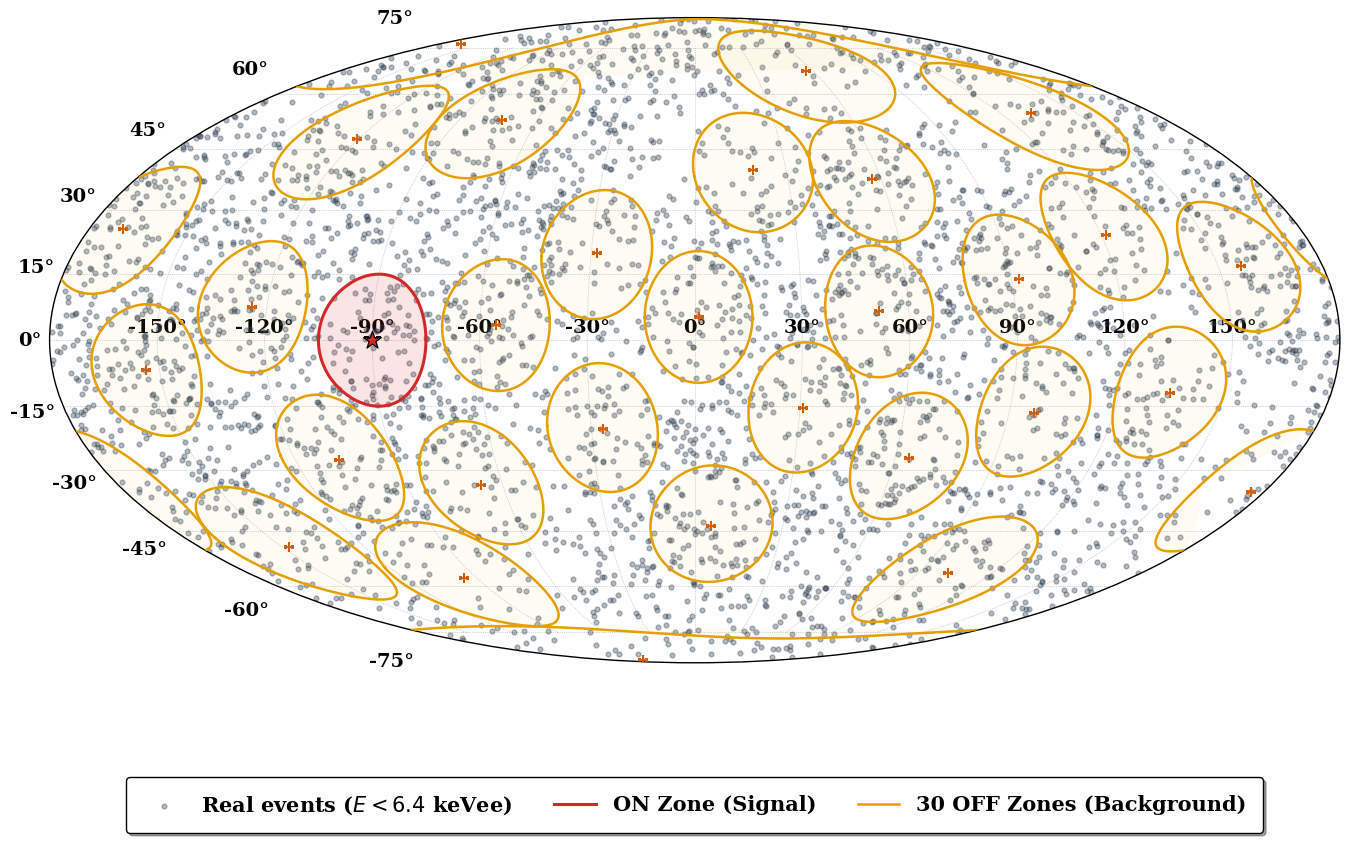} 
    \caption{Galactic map of nuclear recoils, corresponding to an energy selection of $E < 6.4$~keVee for a $10$~GeV/$c^2$ WIMP mass search. The red circle defines the ON-zone ($15^\circ$ radius) centered on the Cygnus constellation. The yellow ellipses represent one iteration of the $N = 30$ OFF-zones used to evaluate the background \textit{in-situ}. The apparent elliptical shapes near the poles and edges are standard geometric distortions induced by the 2D equal-area projection of perfect 3D cones.}    \label{fig:on_off_map}
\end{figure}

In the Standard Halo Model, the dark matter halo of the Milky Way is assumed to be an isothermal, isotropic sphere with no net rotation. Because this halo is globally static, the rapid orbital motion of the Solar System through it ($v_{\odot} \approx 220 \text{ km/s}$) creates a strong, highly anisotropic dipole in the WIMP flux, commonly referred to as the WIMP "wind". This apparent wind originates from the direction of the Cygnus constellation ($l \approx 90^\circ, b \approx 0^\circ$) \cite{REF_Cygnus_Wind_Theory}. Consequently, WIMP-induced nuclear recoils are expected to point predominantly in the opposite direction, towards the Anti-Cygnus constellation ($l \approx -90^\circ, b \approx 0^\circ$). We define our Signal region (ON-zone) as a cone of $15^\circ$ radius centered directly on this expected recoil direction (Anti-Cygnus, $l \approx -90^\circ, b \approx 0^\circ$), see Fig.~\ref{fig:on_off_map}. This specific value is optimized to enclose the maximum signal acceptance by accounting for both the intrinsic kinematic angular dispersion of WIMP-nucleon elastic scattering and the measured angular resolution of the MIMAC $\mu$-TPC.
To evaluate the background  without relying on systematic-heavy simulations, we generate $N=30$ OFF-zones of equal solid angle randomly distributed across the sky map (see Fig.~\ref{fig:on_off_map}). Dynamic vetoes prevent these OFF-zones from overlapping with the ON-zone. Measuring the background from the exact same dataset automatically cancels out environmental and time-dependent detector fluctuations.
To maximize the signal-to-noise ratio in our Region of Interest, we optimize the search phase-space by applying a moving kinematic cut. For a given WIMP mass $m_{\chi}$, the maximum nuclear recoil energy $E_{nr}^{max}$ transferred to a target nucleus of mass $m_T$ is strictly bounded by the maximum relative WIMP-Earth velocity $v_{max} = v_c + v_{esc} \approx 764 \text{ km/s}$, given a local circular velocity $v_c \approx 220 \text{ km/s}$ and a Galactic escape velocity $v_{esc} \approx 544 \text{ km/s}$~\cite{Lavalle2015}. This maximum kinetic energy is given by:
\begin{equation}
    E_{nr}^{max} = \frac{2 \mu_{\chi T}^2 v_{max}^2}{m_T}
\end{equation}
where $\mu_{\chi T}$ is the WIMP-nucleus reduced mass and $m_T$ is the nucleus target mass. 

Because the MIMAC detector measures ionization energy ($E_{ee}$) rather than the true nuclear recoil energy, we must convert this theoretical kinematic boundary into an observable threshold using the measured Ionization Quenching Factor (IQF). This yields a dynamic maximum observable energy, $E_{ee}^{max}$, which scales with the tested WIMP mass, as illustrated for a proton target ($^{1}$H) in Table~\ref{tab:moving_cut}.

\begin{table}[htbp]
    \centering
    \begin{tabular}{c c c c}
        \hline\hline
        WIMP Mass &  $E_{nr}^{max}$ & $E_{ee}^{max}$ & Remaining Events \\
        ($\text{GeV}/c^2$) &  ($\text{keV}$) &  ($\text{keVee}$) & ($N_{\text{ON}}$) \\
        \hline
        1.37   & 4.30  & 2.36  & 9 \\
        5.74   & 9.00  & 5.55  & 73 \\
        9.24   & 10.04 & 6.29  & 75 \\
        18.87  & 11.06 & 7.03  & 77 \\ 
        48.94  & 11.73 & 7.51  & 80 \\
        
        \hline\hline
    \end{tabular}
    \caption{Dynamic kinematic thresholds and remaining events applied to the hydrogen recoil search channel. The physical maximum nuclear recoil energy ($E_{nr}^{max}$) is converted into the observable ionization energy ($E_{ee}^{max}$) using the measured quenching factor, providing a moving upper cut for the ROI and dynamically reducing the background event count.}
    \label{tab:moving_cut}
\end{table}

Instead of using a fixed energy window, our energy cut slides dynamically. This strict kinematic filtering systematically rejects high-energy background events that are physically impossible for lighter WIMPs to produce, drastically improving the signal-to-background ratio at low masses.

The statistical extraction relies on a joint Poisson likelihood framework, coupling the observed number of events in the ON region ($N_{\text{on}} \sim \text{Poisson}(S + \alpha B)$) with the total observed events in the OFF regions ($N_{\text{off}} \sim \text{Poisson}(B)$). Here, $S$ and $B$ represent the expected signal and background events, respectively, while $\alpha = 1/N = 1/30$ is the solid angle ratio between the ON-zone and the $N=30$ OFF-zones. This scaling assumes a spatially isotropic background across the galactic map. Finding no statistically significant excess of events in the signal region above the expected data-driven background, we proceed to set an exclusion limit. We extract the 90\% Confidence Level (CL) upper limit on the number of signal events by computing the Profile Likelihood Ratio. To guarantee absolute statistical robustness and prevent biases from a localized background fluctuation in a single static OFF-zone placement, we implement a spatial bootstrapping technique. The random placement of the 30 OFF-zones is re-sampled 300 times. The final 90\% CL upper limit is defined as the median of these 300 extracted limits, completely immunizing the result against local spatial background anomalies.
Ultimately, the remarkable sensitivity achieved in this work comes from the combination of our topological 3D tracks, ionization quenching factors, and directional observables. In contrast to conventional direct detection experiments that only rely on energy deposition and background simulations, our approach exploits the full 3D track geometry and its distinct morphological properties. By simultaneously combining 3D track reconstruction, calibrated ionization energy, and advanced BDT topological variables, we are able to identify and filter out background events. When this powerful track-based rejection is coupled with the dynamic, mass-dependent kinematic energy cut (the moving cut) and the data-driven spatial ON/OFF analysis, we achieve a  clean signal window, where the electron recoil background is eliminated by the BDT, while the remaining nuclear recoil background is controlled and estimated by using our multiple OFF-zones. This approach is what allows the low-pressure MIMAC TPC without any shielding to establish highly competitive limits on the WIMP-proton spin dependent cross section with a very modest target mass.
To extract a physical cross-section limit, we adopt a local dark matter density of $\rho_0 = 0.44\text{ GeV/cm}^3$, in agreement with recent stellar kinematics determinations based on Gaia Data Release 3 (DR3) measurements~\cite{Soding:2025local}. The active volume of each MIMAC chamber is $10.8 \times 10.8 \times 25 \text{ cm}^3$. Operating at 30 mbar with a room-temperature (293 K) gas mixture of 50\% $\text{i-C}_4\text{H}_{10}$ and 50\% $\text{CHF}_3$, the total active target mass is strictly determined using the ideal gas law. Accounting for the 10 hydrogen atoms in isobutane and the single hydrogen atom in $\text{CHF}_3$, the effective hydrogen target mass is approximately 19.9 mg per chamber. Combining the operational periods of 495.1 days for ch0 and 354.2 days for ch1, we achieve a total effective hydrogen exposure of  $\approx 0.02 \text{ kg}\cdot\text{days}$. 
\begin{table}[htbp]
    \centering
    \begin{tabular}{l c}
        \hline\hline
        Source of Uncertainty & Rel. Error on Acceptance \\
        \hline
        Ionization Quenching Factor (IQF) & 2.0 \% \\
        Charge Drift Velocity (MIMAC)\cite{Billard2014_Drift}  & 1.0 \% \\
        BDT Classifier Efficiency         & 11.6 \% \\
        Gas Pressure ($30.0 \pm 0.3$ mbar)    & 1.0 \% \\
        Gas Temperature ($\Delta T/T$)    & 1.0 \% \\
        \hline
        \textbf{Total Systematic Error ($\delta_{sys}$)} & \textbf{11.9\%} \\
        \hline\hline
    \end{tabular}
    \caption{Summary of the relative systematic uncertainties on the signal acceptance. Astrophysical, instrumental, and analysis errors are added in quadrature to extract a conservative upper limit.}
    \label{tab:systematics}
\end{table}
While this exposure is orders of magnitude smaller than those of monolithic non-directional detectors, 3D track directionality acts as a profound statistical multiplier. Phenomenological studies establish that directional detection provides at least an order of magnitude gain in sensitivity, particularly for low-mass WIMPs, by exploiting the strong dipole signature of the galactic wind against isotropic low energy neutron backgrounds \cite{REF_Directional_Review}. Furthermore, mathematical frameworks demonstrate that integrating angular information (e.g., via 2D spatial methods \cite{REF_Henderson_MaximumPatch}) systematically outperforms 1D energy-only limit-setting techniques, such as the standard Maximum Gap method \cite{REF_Yellin_MaximumGap}. This multidimensional approach produces significantly tighter exclusion bounds for equivalent datasets. Consequently, a directional detector operating in an in-situ background-controlled regime—achieved here via strict topological and spatial ON/OFF filtering—can extract an unambiguous signal from the background with a minimal number of events, yielding highly competitive exclusion limits despite a low target mass \cite{REF_Billard_Discovery}.

The total systematic uncertainty on the global signal acceptance ($\Delta \epsilon / \epsilon$) is evaluated by summing the independent relative errors in quadrature. The individual contributions from the halo kinematics, ionization quenching, detector conditions, and selection efficiencies are detailed in Table~\ref{tab:systematics}.
\begin{figure}[htbp]
    \centering

    \includegraphics[width=1.0\columnwidth]{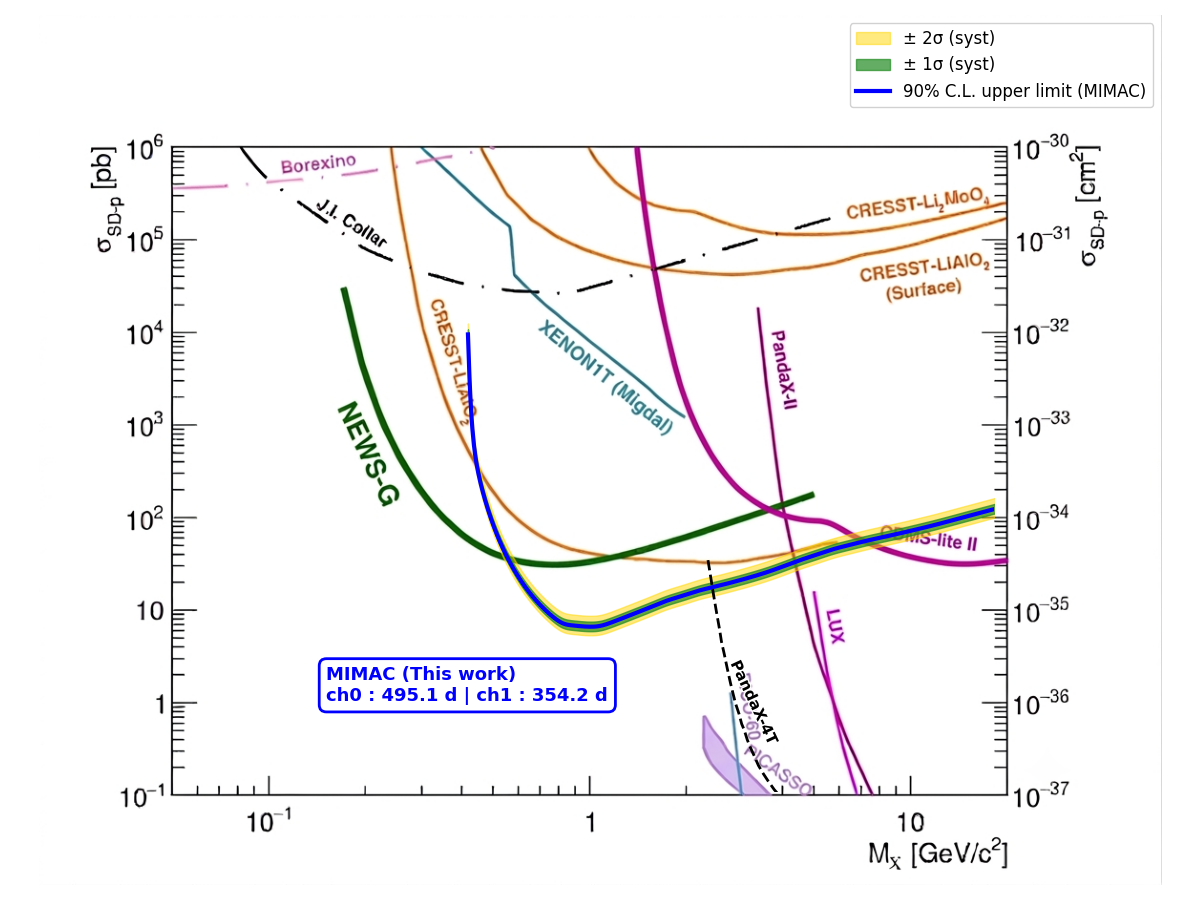} 
     \caption{The 90\% CL upper limit on the spin-dependent WIMP-proton cross section ($\sigma_{SD-p}$) obtained from the combined MIMAC ch0 and ch1 data (blue solid line). The inner (green) and outer (yellow) shaded bands represent the $\pm 1\sigma$ and $\pm 2\sigma$ total propagated systematic uncertainties, respectively. For comparison, we display the experimental limits set by leading direct detection searches: the sub-GeV methane-target search of NEWS-G~\cite{REF_Arnaud_NEWSG}, the cryogenic germanium-based CDMSlite~\cite{REF_Agnese_CDMSlite}, the low-threshold lithium and aluminate-based CRESST programs~\cite{REF_Abdelhameed_CRESST, REF_Angloher_CRESST_6Li, REF_Angloher_CRESST_LiAlO2}, the complete exposure limits from LUX~\cite{REF_Akerib_LUX} and PandaX-II~\cite{REF_Xia_PandaX_II}, the Migdal-enhanced search of XENON1T~\cite{REF_Aprile_XENON}, the PICASSO bubble chamber~\cite{REF_Behnke_PICASSO}, the PICO-60 complete dataset~\cite{REF_Amole_PICO60}, and the recent sub-GeV search results from PandaX-4T~\cite{REF_Lin_PandaX_4T}.}
    \label{fig:exclusion_curve}
\end{figure}

Astrophysical uncertainties affect the moving kinematic cut, while instrumental fluctuations (such as $P$ and $T$) directly impact the gas density and the total target mass. The uncertainties in the drift velocity, evaluated from in-situ calibrations of the detector drift parameters~\cite{Billard2014_Drift}, and the efficiency of the BDT affect the reconstruction and selection of 3D tracks. These errors are combined in quadrature to give a total relative systematic uncertainty $\delta_{sys}$. The statistical 90\% CL upper limit extracted from the Likelihood is reported as the nominal limit. The total systematic uncertainty is then analytically propagated to construct $\pm 1\sigma$ and $\pm 2\sigma$ uncertainty bands around this nominal limit. This standard representation transparently illustrates the robustness of the directional exclusion bounds against both instrumental and astrophysical fluctuations.

Following this procedure, no significant excess of nuclear recoil events was observed in the Cygnus direction above the data-driven background estimation. Fig.~\ref{fig:exclusion_curve} displays the resulting 90\% CL upper limit on the spin-dependent WIMP-proton cross section as a function of the WIMP mass.

In conclusion, the MIMAC $\mu$-TPC detector demonstrates that 3D track directionality combined with topological discrimination provides a performant, active background rejection capability. Using a purely data-driven spatial analysis, we securely bypassed the uncertainties of simulated backgrounds. Reaching a limit of $\sigma_{SD-p} <6.65 \times 10^{-36} \text{ cm}^2$ without massive passive shielding highlights the potential of directional detection. Future campaigns with the upgraded MIMAC-35cm detector (twelve times larger active volume and 100 eV energy threshold) will allow us to probe the low-mass WIMP parameter space with unprecedented directional sensitivity.

\begin{acknowledgments}
We thank the Modane Underground Laboratory and its staff for support through underground space, logistical and technical services. LSM operations are supported by the CNRS, with underground access facilitated by the Société Française du Tunnel Routier du Fréjus. This work is supported by the Labex Enigmass.

\end{acknowledgments}

\nocite{*}
\bibliography{mimac} 

\end{document}